\documentclass[final,3p,times]{elsarticle}

\usepackage{amssymb}
\usepackage{amsmath}
\usepackage{color}
\usepackage{url}

\journal{Journal of Subatomic Particles and Cosmology}

\begin{document}

\begin{frontmatter}

\title{Microscopic Insights into the Quarkyonic Hadron--Quark Crossover: \\
Lessons from Ultracold Fermi Gases}

\author[utokyo,riken,qnsi]{Hiroyuki Tajima}

\affiliation[utokyo]{organization={Department of Physics, The University of Tokyo},
                     city={Tokyo},
                     country={Japan}}
\affiliation[riken]{organization={RIKEN Nishina Center},
                     city={Wako},
                     country={Japan}}
\affiliation[qnsi]{organization={Quark Nuclear Science Institute, The University of Tokyo},
                     city={Tokyo},
                     country={Japan}}

\begin{abstract}
In neutron-star and dense quantum chromodynamics (QCD) physics,
it is important to understand how baryonic matter evolves into quark matter at high density.
A continuous hadron--quark crossover is one of the most attractive candidates
because it can reconcile the
comparatively soft equation of state around nuclear density with the rapid
stiffening required to support massive neutron stars. 
Recent neutron-star observations motivate equations of state that stiffen
rapidly across the intermediate-density regime, while quarkyonic-matter
descriptions feature a shell-like baryon momentum distribution there.
In this paper,
we review our recent work on the microscopic description and interpretation of these phenomena by drawing an analogy with the
Bose--Einstein condensate (BEC) to Bardeen--Cooper--Schrieffer (BCS) crossover in ultracold Fermi gases and two-color QCD.  
As pairing fluctuations play a crucial role in the BEC--BCS crossover, 
we discuss the role of its three-body counterpart, that is, tripling fluctuations associated with baryon formation in the hadron--quark crossover. 
In terms of a phase-shift representation of tripling fluctuations, the interplay between a three-body bound-state
pole and the scattering continuum suppresses low-momentum baryonic
occupation, generates a baryonic momentum shell, and reduces the density
susceptibility, thereby producing a peak in the speed of sound.  
We review
the demonstration of this mechanism in a one-dimensional three-component
Fermi gas, its relativistic extension, and its connection to phenomenological
quarkyonic equations of state.  
\end{abstract}


\end{frontmatter}

\section{Introduction}
\label{sec:introduction}

The composition of strongly interacting matter at baryon densities above the
nuclear saturation density is one of the most important problems in nuclear astrophysics and in dense quantum chromodynamics (QCD) physics.
Microscopic nuclear
calculations constrain the equation of state at relatively low density,
whereas quark matter descriptions are expected to
become appropriate at asymptotically high density.  Neutron stars probe the
intermediate region in which neither description is manifestly complete.
The existence of neutron stars with masses near two solar masses requires a
sufficiently stiff equation of state~\cite{demorest2010two,antoniadis2013massive,cromartie2020relativistic}, while radius and tidal-deformability
measurements constrain how this stiffening can occur~\cite{PhysRevLett.119.161101,PhysRevX.9.011001}.

A continuous hadron--quark crossover (HQC) provides one possible explanation of these observations~\cite{Baym2018}.  In such a scenario,
the relevant degrees of freedom continuously change from baryons to quarks without phase transitions.
Moreover, quarkyonic
matter description~\cite{mclerran2007phases} offers a useful realization of the HQC scenario, where deeply occupied momentum
states are described predominantly by quarks and baryonic excitations
remain near the Fermi surface.  
Indeed, phenomenological quarkyonic models
predict both rapid stiffening and a nonmonotonic speed of sound~\cite{McLerranReddy2019}.  
The quarkyonic matter model has been further elaborated within the variational approach~\cite{Fujimoto2024}.
So far, extensions of quarkyonic matter to various models such as 
relativistic mean-field model~\cite{PhysRevD.108.054013},
parity-doublet model~\cite{PhysRevD.111.016024}, 
and quark-meson coupling model~\cite{s71s-85r5}, 
have also been discussed.

From the viewpoint of lattice simulations,
two-color QCD provides a particularly transparent connection between dense
gauge theory and crossover physics~\cite{suenaga2025chiral}.
Because the fundamental representation
of $SU(2)$ is pseudoreal, color-singlet diquarks can be bosonic baryons.  Dense
matter may therefore evolve from a Bose--Einstein condensate (BEC) of tightly
bound diquarks to a Bardeen-Cooper-Schrieffer (BCS) regime (i.e., BEC--BCS crossover~\cite{strinati2018bcs,Ohashi2020}) in which quarks
pair near a Fermi surface~\cite{KOGUT2000477,IidaItouLee2020}.
Indeed, lattice simulations find an enhanced sound speed across this crossover
\cite{IidaItou2022,IidaEtAl2024}, making this theory a useful setting for
examining the connection between composite formation and a stiff equation
of state.

The BEC--BCS crossover has been realized in ultracold Fermi gases
\cite{PhysRevLett.92.040403,PhysRevLett.92.120403,chin2004observation}.
Two-color QCD makes the crossover analogy concrete, but ordinary
three-color baryons are three-fermion composites rather than pairs.
This review takes the fluctuation theory of ultracold Fermi gases as a
starting point for a microscopic account of these signatures.  In the
BEC--BCS crossover, pairing fluctuations describe how two-fermion bound
states evolve into correlations near a Fermi surface.  For three-color
fermions the analogous channel is instead fermionic: three-body correlations
can persist as Cooper triples at finite density
\cite{PhysRevA.86.013628,PhysRevA.104.053328,Tajima2022}.
The tripling-fluctuation approach relates the baryonic momentum shell and
the enhanced sound speed to the interplay between a three-body bound-state
pole and the scattering continuum~\cite{Tajima2025,Tajima2026}.
We review
this mechanism in a one-dimensional model and compare it with phenomenological
quarkyonic constructions.

In the following, we set $\hbar=k_B=1$ and take unit volume.

\section{Pairing fluctuations in the BEC--BCS crossover}
\label{sec:becbcs}
In this section, we review pairing fluctuations in the BEC--BCS crossover.
The Nozi\`eres--Schmitt-Rink (NSR) theory incorporates pairing fluctuations
in the normal state of an attractive Fermi gas~\cite{NSR1985,PhysRevLett.71.3202}.
We consider a uniform balanced two-component Fermi gas in three dimensions,
described by the grand-canonical Hamiltonian density
\begin{equation}
 \mathcal H_2=\sum_{\sigma=\uparrow,\downarrow}
 \psi_\sigma^\dagger\left(-\frac{\nabla^2}{2m}-\mu\right)\psi_\sigma
 +V_2\psi_\uparrow^\dagger\psi_\downarrow^\dagger
 \psi_\downarrow\psi_\uparrow.
 \label{eq:pair_hamiltonian}
\end{equation}
Here $\psi_\sigma$ annihilates a fermion in component $\sigma$, $m$ is the
common mass, $\mu$ is the chemical potential of each component, and
$V_2<0$ is the two-body contact coupling.  
The free dispersion is $\xi_{\bf p}=\varepsilon_{\bf p}-\mu$, with
$\varepsilon_{\bf p}=p^2/(2m)$ and $n_{\rm F}(x)=(e^{x/T}+1)^{-1}$.

Pairing fluctuations in this system are described by the in-medium $T$ matrix
\begin{align}
 \Gamma_2({\bf K},z)&=\frac{V_2}{1-V_2G_{2,0}({\bf K},z)},
 \label{eq:nsr_tmatrix}\\
 G_{2,0}({\bf K},z)&=\sum_{\bf p}
 \frac{1-n_{\rm F}(\xi_+)-n_{\rm F}(\xi_-)}{z-\xi_+-\xi_-},
 \label{eq:nsr_bubble}
\end{align}
where $\xi_\pm=\xi_{\pm{\bf p}+{\bf K}/2}$ and
$z=i\nu_n$ with the bosonic Matsubara frequency $\nu_n=2\pi nT$.
The bare pair propagator $G_{2,0}$ is built from two free single-particle
propagators at finite $T$ and $\mu$.

The pairing-fluctuation correction to the thermodynamic potential is given by 
\begin{align}
 \delta\Omega_2&=T\sum_{{\bf K},i\nu_n}
 \ln[1-V_2G_{2,0}({\bf K},i\nu_n)],
 \label{eq:nsr_omega}
\end{align}
To express the pairing fluctuations in terms of a phase shift, we introduce an absolute
two-particle energy $\omega$, such that analytic continuation is
$z\to\omega-2\mu+i0^+$, and define the two-body scattering phase shift~\cite{PhysRev.187.345}
\begin{equation}
 \phi_2({\bf K},\omega)
 =-\operatorname{Im}\ln[1-V_2G_{2,0}({\bf K},\omega-2\mu+i0^+)].
 \label{eq:nsr_phase}
\end{equation}
Contour deformation of the Matsubara summation in Eq.~\eqref{eq:nsr_omega} then gives
\begin{equation}
 \delta\Omega_2=-\sum_{\bf K}\int_{-\infty}^{\infty}
 \frac{d\omega}{\pi}\,n_{\rm B}(\omega-2\mu)\phi_2({\bf K},\omega).
 \label{eq:nsr_phase_omega}
\end{equation}
The Bose distribution function $n_{\rm B}(x)=(e^{x/T}-1)^{-1}$ in Eq.~\eqref{eq:nsr_phase_omega} reflects the statistics of a pair of fermions.  This
representation incorporates both bound pairs and scattering states through
the energy dependence of the same phase shift.

\section{Tripling fluctuations in the three-body crossover}
\label{sec:tripling}
To demonstrate the three-body counterpart of pairing fluctuations, that is, tripling fluctuations, we consider a nonrelativistic three-component Fermi gas with an attractive three-body
contact interaction~\cite{Drut2018,PhysRevA.100.063604,PhysRevC.109.055203,hryhorchak2025bipolaron} described by the Hamiltonian density
\begin{align}
 {\cal H}_3={}&\sum_{a=r,g,b}\psi_a^\dagger
 \left(-\frac{\nabla^2}{2m}-\mu\right)\psi_a+V_3(\psi_r^\dagger\psi_r)
 (\psi_g^\dagger\psi_g)(\psi_b^\dagger\psi_b),
 \label{eq:1d_hamiltonian}
\end{align}
with an attractive three-body coupling $V_3<0$.  Regularization of this classically scale-invariant coupling $V_3$
introduces the vacuum binding energy $\mathcal B$ and a nonrelativistic
trace anomaly in one spatial dimension~\cite{Drut2018,PhysRevA.100.063604,PhysRevC.109.055203}. 
Moreover, the present model admits
sign-problem-free quantum Monte Carlo benchmarks as performed in Ref.~\cite{PhysRevA.102.023313}.
For completeness, in what follows, we show the tripling-fluctuation formalism in general spatial dimensions.

In analogy with pairing fluctuations, tripling fluctuations can be described by the in-medium three-body $T$ matrix~\cite{Tajima2022}
\begin{equation}
 \Gamma_3({\bf K},z)=\frac{V_3}{1-V_3G_{3,0}({\bf K},z)}.
 \label{eq:three_body_tmatrix}
\end{equation}
The in-medium three-body propagator $G_{3,0}({\bf K},z)$ is given by
\begin{align}
 G_{3,0}({\bf K},z)
 &=\sum_{{\bf k},{\bf q}}
 \frac{(1-f_1)(1-f_2)(1-f_3)+f_1f_2f_3}
 {z-\xi_1-\xi_2-\xi_3},
 \label{eq:g30_medium}
\end{align}
where $f_i=n_{\rm F}(\xi_i)$, $\xi_i=p_i^2/(2m)-\mu$, $z=i\omega_n$, and the constituent momenta are
 ${\bf p}_1=\frac{{\bf K}}{3}+{\bf k}-\frac{{\bf q}}{2}$,
 ${\bf p}_2=\frac{{\bf K}}{3}+{\bf q}$, and
 ${\bf p}_3=\frac{{\bf K}}{3}-{\bf k}-\frac{{\bf q}}{2}$.

Tripling-fluctuation correction to the thermodynamic potential
is thus given by
\begin{equation}
 \delta\Omega_3=-T\sum_{{\bf K},i\omega_n}
 \ln[1-V_3G_{3,0}({\bf K},i\omega_n)],
 \label{eq:omega3_matsubara}
\end{equation}
where $\omega_n=(2n+1)\pi T$ is the fermion Matsubara frequency.
In Eq.~\eqref{eq:omega3_matsubara}, relative momenta and frequencies have been summed in $G_{3,0}$.
If a
Hartree--Fock contribution is included separately, its first-order term
must be removed from the fluctuation term as in Ref.~\cite{Tajima2025}.
For Eq.~(\ref{eq:omega3_matsubara}), analytic continuation to the absolute
three-particle energy and contour deformation give
\begin{equation}
 \delta\Omega_3=-\sum_{\bf K}\int_{-\infty}^{\infty}
 \frac{d\omega}{\pi}\,n_{\rm F}(\omega-3\mu)\phi_3({\bf K},\omega),
 \label{eq:omega3}
\end{equation}
where
$\phi_3({\bf K},\omega)=-\operatorname{Im}\ln[1-V_3G_{3,0}^{\rm R}({\bf K},\omega)]
$ is the three-body scattering phase shift~\cite{PhysRev.187.345}.
The superscript $R$ of $G_{3,0}^{\rm R}({\bf K},\omega)$ denotes retarded propagators evaluated with
$i\omega_n\to\omega-3\mu+i0^+$.
Comparison with Eq.~(\ref{eq:nsr_phase_omega}) makes the change from
bosonic pairs to fermionic triples explicit: the statistical weight is
now a Fermi distribution function $n_{\rm F}(\omega-3\mu)$ with the chemical potential of three constituents.

For the analytically tractable contact model considered below, it is useful
to define the cluster spectral-weight difference entering the density by
the derivative of the phase shift,
\begin{equation}
 \Delta A_3({\bf K},\omega)
 =\frac{1}{\pi}\frac{\partial\phi_3({\bf K},\omega)}{\partial\omega}.
 \label{eq:phase_spectral}
\end{equation}
This definition refers to the spectral contribution in the phase-shift
representation.
With the phase chosen to vanish
at both frequency endpoints, it obeys
\begin{equation}
 \int_{-\infty}^{\infty}d\omega\,\Delta A_3({\bf K},\omega)
 =\frac{\phi_3({\bf K},\infty)-\phi_3({\bf K},-\infty)}{\pi}=0.
 \label{eq:sumrule}
\end{equation}
Accordingly, a positive delta-function contribution from a bound-state pole
is accompanied by a negative continuum contribution.  Keeping only the pole
misses the cancellation that becomes decisive for the baryon-like momentum
distribution~\cite{Tajima2025,Tajima2026}.

Differentiating $\delta\Omega_3$ with respect to the constituent chemical
potential yields the fluctuation contribution to the constituent number
density.  When the phase at fixed absolute energy has no additional
chemical-potential dependence, integration by parts gives an effective
cluster distribution $f_B({\bf K})$ through
\begin{align}
 n_{\mathrm{fluc}}=3\sum_{\bf K}f_B({\bf K}),\qquad
 f_B({\bf K})=\int_{-\infty}^{\infty}d\omega\,
 n_{\rm F}(\omega-3\mu)\Delta A_3({\bf K},\omega).
 \label{eq:cluster_density}
\end{align}
The factor three in Eq.~\eqref{eq:cluster_density} counts the fermionic
constituents; $n_{\mathrm{fluc}}/3$
is the baryon-like contribution to the baryon number density.

\section{Three-body crossover at low temperatures}
\label{sec:crossover}
To examine the three-body crossover in the one-dimensional three-color fermions described by Eq.~\eqref{eq:1d_hamiltonian},
we discuss the evolution of the three-body pole in the three-body $T$ matrix $\Gamma_3({K},z)$~\cite{Tajima2022}.
Analytic continuation $z\to\omega-3\mu+i0^+$ defines
$G_{3,0}^{\rm R}(K,\omega)$ and $\Gamma_3^{\rm R}(K,\omega)$ at absolute energy $\omega$.
At $K=0$, the free three-particle continuum starts at $\omega=0$,
and the in-medium binding energy $\mathcal B_{\rm M}>0$ obeys
\begin{equation}
 [\Gamma_3^{\rm R}(0,-\mathcal B_{\rm M})]^{-1}
 =\frac{1}{V_3}-G_{3,0}^{\rm R}(0,-\mathcal B_{\rm M})=0.
 \label{eq:trimer_medium_pole}
\end{equation}
In the frequency measured relative to $3\mu$, the same pole lies at
$z=-\mathcal B_{\rm M}-3\mu$.  
We note that the three-body binding energy $\mathcal{B}=\frac{\Lambda^2}{m}e^{\frac{2\sqrt{3}\pi}{mV_3}}$ is obtained in the dilute limit ($T\rightarrow 0$ and $\mu\rightarrow-\infty$) of Eq.~\eqref{eq:trimer_medium_pole}, where $\Lambda$ is the momentum cutoff.

\begin{figure}[t]
    \centering
    \includegraphics[width=1\linewidth]{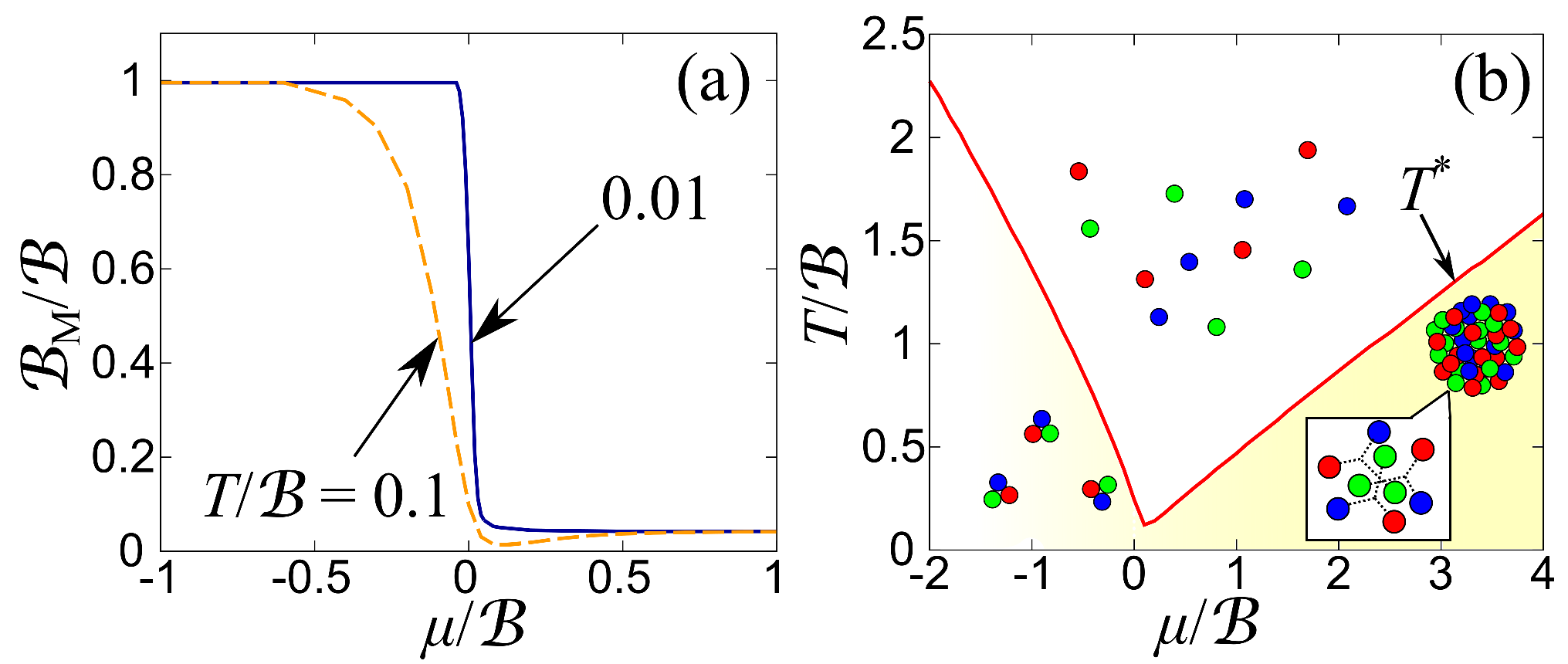}
    \caption{(a) In-medium binding energy $\mathcal B_{\rm M}$ at $T/\mathcal B=0.01$ and $0.1$. 
    (b) Schematic grand-canonical diagram: the line $T^*$ marks the Mott temperature where the in-medium three-body bound-state energy vanishes.
    Adapted with modified labels and colors from Figs. 2 and 4 of H. Tajima, S. Tsutsui, T. M. Doi, and K. Iida, ``Cooper triples in attractive three-component fermions: Implication for hadron-quark crossover,'' Phys. Rev. Research 4, L012021 (2022)~\cite{Tajima2022}, \protect\url{https://doi.org/10.1103/PhysRevResearch.4.L012021}, licensed under CC BY 4.0 (\protect\url{https://creativecommons.org/licenses/by/4.0/}).}
    \label{fig:1}
\end{figure}

Figure~\ref{fig:1}(a) shows the numerical result of  
$\mathcal B_{\rm M}$~\cite{Tajima2022},
which drops near $\mu=0$
but remains finite at higher density for both temperatures shown.
This result indicates that
in-medium three-body correlations survive even under the formation of a Fermi sea.
This is reminiscent of the quarkyonic matter picture~\cite{mclerran2007phases}, where baryon degrees of freedom survive even as a quark Fermi sea forms.

Figure~\ref{fig:1}(b) shows the finite-density phase diagram of a one-dimensional three-color Fermi gas at finite temperature.
The Mott temperature $T^*$ where the bound-state pole disappears (i.e., $\mathcal{B}_{\rm M}\rightarrow 0$)~\cite{strinati2018bcs} shows the characteristic scale for three-body correlations.
While $T^*$ does not indicate a thermodynamic phase boundary,
$T^*$ does not go to zero, and hence the crossover with strong three-body correlations at sufficiently low temperature can be expected.

\section{Baryonic momentum-shell formation}
\label{sec:shell}
\begin{figure}
    \centering
    \includegraphics[width=\linewidth]{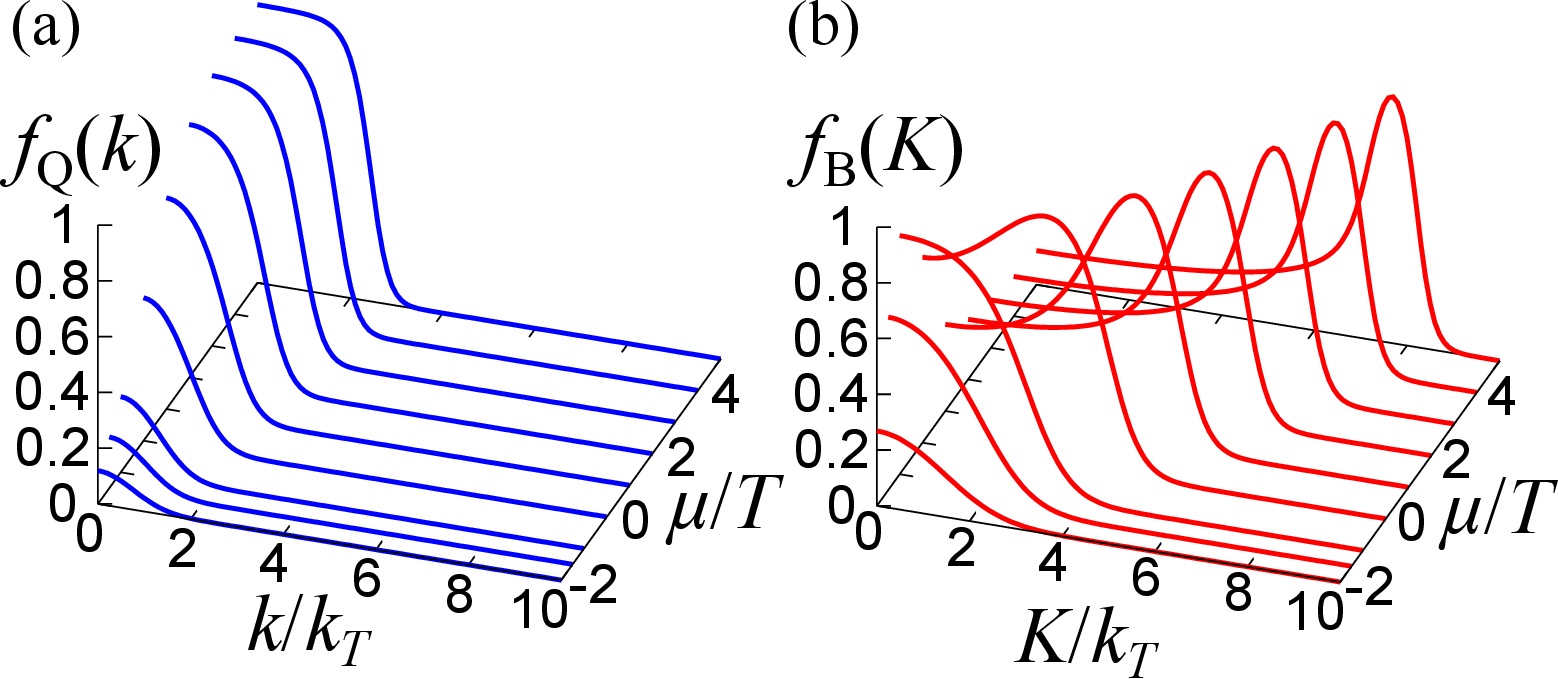}
    \caption{Momentum distributions of (a) quark-like constituents and (b) baryon-like three-body correlations at $T/\mathcal B=0.2$, with $k_T=\sqrt{2mT}$. Reproduced from Fig. 2 of H. Tajima, K. Iida, T. Kojo, and H. Liang, ``Tripling Fluctuations and Peaked Sound Speed in Fermionic Matter,'' Phys. Rev. Lett. 135, 042701 (2025)~\cite{Tajima2025}, \protect\url{https://doi.org/10.1103/4ywp-752m}, licensed under CC BY 4.0 (\protect\url{https://creativecommons.org/licenses/by/4.0/}).}
    \label{fig:2}
\end{figure}

To analyze the pole--continuum cancellation analytically, we set
$f_i\to0$ inside $G_{3,0}$, while retaining the cluster Fermi
weights in Eq.~(\ref{eq:cluster_density})~\cite{Tajima2025}.
This approximation is
valid for large $K$, a regime relevant to the baryon momentum-shell formation.
Figure~\ref{fig:2}(a) and (b) compare the constituent distribution $f_Q(k)=n_{\rm F}(k^2/2m-\tilde{\mu})$ with
$f_B(K)$ at $T/\mathcal B=0.2$, using a thermal momentum scale $k_T=\sqrt{2mT}$ (where $\tilde\mu=\mu-\Sigma_{\rm HF}$ is the chemical potential with the Hartree-Fock self-energy $\Sigma_{\rm HF}$).
As $\mu/T$ increases, $f_Q$ approaches a Fermi step, whereas $f_B$
is depleted near $K=0$ and peaks at finite $|K|$.  

The mechanism is transparent in the approximation of
Ref.~\cite{Tajima2025}, for which
\begin{align}
 f_B(K)={}&n_{\rm F}\!\left(\frac{K^2}{2M_B}-3\tilde\mu-\mathcal B\right)
 -\int_0^\infty\!d\omega\,
 \frac{n_{\rm F}\!\left(\omega+\frac{K^2}{2M_B}-3\tilde\mu\right)}
 {\omega[\ln^2(\omega/\mathcal B)+\pi^2]},
 \label{eq:fB_pole_continuum}
\end{align}
where $M_B=3m$.
The first term in Eq.~\eqref{eq:fB_pole_continuum} is the bound-state pole and the second one is the negative
scattering-continuum contribution.  Their cancellation at small
$K$ produces the depletion in Fig.~\ref{fig:2}(b).
In the dilute regime, the pole dominates because the chemical potential
lies below the continuum.  With increasing density, occupied scattering
states cancel the pole at small $K$, while at larger $K$ the continuum
can be unoccupied and the pole still occupied.

At $T=0$ the continuum ceases to contribute above the threshold
$K_{\rm th}=3k_{\rm F}$, with $k_{\rm F}=\sqrt{2m\tilde\mu}$, and the shell extends
approximately from $K_{\rm th}$ to
$\sqrt{K_{\rm th}^2+2M_B\mathcal B}$ in this model
\cite{Tajima2025}.  Its width reflects the binding scale and shrinks as
$k_{\rm F}$ increases; finite temperature smooths its edges.
The shell here follows from the pole--continuum spectrum and its Fermi
weights.  This complements the phenomenological shell width in the
McLerran--Reddy model~\cite{McLerranReddy2019} and the explicit
quark--baryon duality constraint in the IdylliQ model~\cite{Fujimoto2024}.

\section{Peaked speed of sound}
\label{sec:soundspeed}
\begin{figure}[t]
    \centering
    \includegraphics[width=0.5\linewidth]{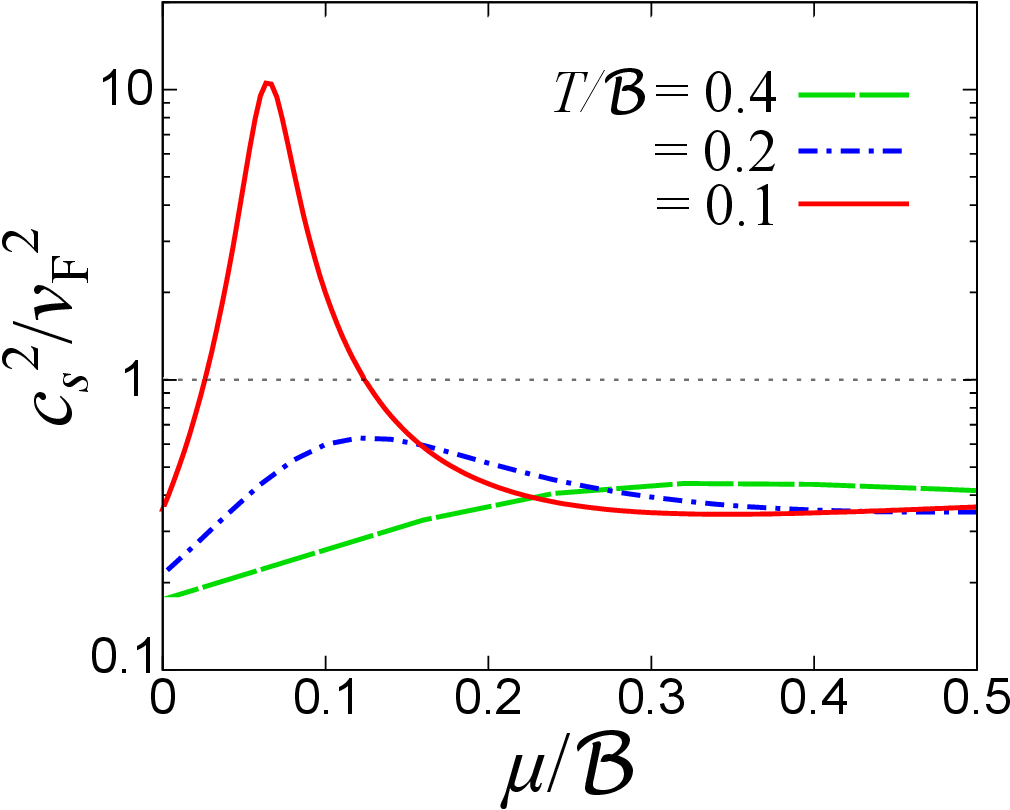}
    \caption{Squared isothermal speed of sound $c_s^2$ normalized by $v_F^2$, where $v_F=\pi n/(3m)$; the dotted line marks the high-density limit. Reproduced from Fig. 3 of H. Tajima, K. Iida, T. Kojo, and H. Liang, ``Tripling Fluctuations and Peaked Sound Speed in Fermionic Matter,'' Phys. Rev. Lett. 135, 042701 (2025)~\cite{Tajima2025}, \protect\url{https://doi.org/10.1103/4ywp-752m}, licensed under CC BY 4.0 (\protect\url{https://creativecommons.org/licenses/by/4.0/}).}
    \label{fig:3}
\end{figure}

In this section,
we show that the peaked speed of sound can be microscopically explained by the appearance of the baryonic momentum shell shown in Fig.~\ref{fig:2}(b).
For the isothermal response of the nonrelativistic model, the squared sound
velocity is related to the density susceptibility by
\begin{equation}
 c_s^2 = \frac{n}{m}\left(\frac{\partial n}{\partial\mu}\right)_T^{-1},
 \label{eq:nonrel_sound}
\end{equation}
where $n$ is the total constituent density and $m$ the constituent mass.
Figure~\ref{fig:3} plots $c_s^2/v_{\rm F}^2$ against $\mu/\mathcal B$ for
three temperatures, where $v_{\rm F}=\pi n/(3m)$ is the Fermi velocity formed
from the total constituent density~\cite{Tajima2025}.  
The dotted line at unity is the
high-density limit of this nonrelativistic model.
The low-temperature curve shows a pronounced maximum
in the crossover region; heating broadens and lowers the peaked structure~\cite{Tajima2025}.

To relate the sound-speed peak to shell formation, we write
$n=n_{\rm HF}+n_{\rm fluc}$.  Then, the density susceptibility reads
\begin{equation}
 \frac{\partial n}{\partial\mu}
 =\frac{\partial n_{\rm HF}}{\partial\mu}
 +\frac{\partial n_{\rm fluc}}{\partial\mu}.
 \label{eq:susceptibility}
\end{equation}
Depletion of $f_B(K)$ at small $K$ can make
$\partial n_{\rm fluc}/\partial\mu$ negative in the crossover region.
In this regard,
$\partial n_{\rm fluc}/\partial\mu$ partially cancels the positive Hartree--Fock response $\partial n_{\rm HF}/\partial\mu$, while stability
requires the total susceptibility to remain positive.  Equation~(\ref{eq:nonrel_sound})
then gives an enhanced $c_s^2$.
In this way, the sound-speed peak and momentum shell
share the same pole--continuum cancellation mechanism~\cite{Tajima2025}.

At zero temperature, the relativistic counterpart is
 $c_{s,{\rm R}}^2=\frac{dP}{d\varepsilon}
 =\frac{n_B}{\mu_B}\left(\frac{dn_B}{d\mu_B}\right)^{-1}$,
where $n_B$ is the total baryon density and $\mu_B=N_c\mu$.
Suppression of the total density susceptibility can therefore enhance the
relativistic sound speed as well.  
While the hydrodynamic
response $(\partial P/\partial\varepsilon)_{s/n_B}$ generally differs
from the isothermal response at finite temperatures,
such a deviation could be negligible at sufficiently low temperatures.

\section{From the phase-shift representation to the McLerran--Reddy model}
\label{sec:relativistic}
Finally, we discuss the relationship between tripling fluctuations and the phenomenological quarkyonic matter picture.
We now follow the construction in Sec.~3.2 of Ref.~\cite{Tajima2026}
to recover the structure of quark-sea plus baryon shell used in the McLerran--Reddy model~\cite{McLerranReddy2019}.  We consider symmetric
two-flavor matter in three spatial dimensions, with $N_c=3$ and a common
spin--isospin degeneracy $g=4$.  Quarks have dispersion
$E_Q(k)=\sqrt{k^2+m_Q^2}$, while the baryon kinetic energy is approximated
by $K^2/(2M_B)$.  Antiparticles and interactions outside the binding channel
are neglected.  The masses and binding energy are held fixed in the
following derivation.  We retain $\mu_B=N_c\mu$ for the full baryon
chemical potential, consistent with Sec.~\ref{sec:soundspeed}.

Let $\phi_B({\bf K},\omega)$ denote the baryonic phase shift and
$\Delta A_B=\pi^{-1}\partial_\omega\phi_B$ the spectral-weight difference
entering the cluster contribution.  The free-quark term is
$\Omega_Q=-gN_cT\sum_{\bf k}\ln[1+e^{-(E_Q(k)-\mu)/T}]$.
In the fixed-spectrum approximation, the phase-shift term and the
corresponding density are
\begin{align}
 \delta\Omega_B&=-g\sum_{\bf K}\int_{-\infty}^{\infty}
 \frac{d\omega}{\pi}\,n_{\rm F}(\omega-\mu_B)\phi_B({\bf K},\omega),
 \label{eq:rel_phase_omega}\\
 n_B&\equiv-\frac{\partial(\Omega_Q+\delta\Omega_B)}{\partial\mu_B}
 =g\sum_{\bf k}n_{\rm F}(E_Q(k)-\mu)
 +g\sum_{\bf K}\int_{-\infty}^{\infty}d\omega\,
 n_{\rm F}(\omega-\mu_B)\Delta A_B({\bf K},\omega).
 \label{eq:rel_phase_density}
\end{align}
The cluster term follows by integration by parts with vanishing endpoint
phases.  The quark color factor $N_c$ cancels each quark's baryon charge
$1/N_c$, so $n_B$ includes both contributions.  Explicit medium dependence
of the spectrum would require additional chemical-potential derivatives.

The pole--continuum ansatz of Ref.~\cite{Tajima2026} reads
\begin{align}
 E_{\rm th}(K)&=N_c m_Q+\frac{K^2}{2M_B},
 \qquad E_B(K)=E_{\rm th}(K)-\mathcal B,
 \label{eq:rel_threshold}\\
 \Delta A_B({\bf K},\omega)
 &\simeq\delta(\omega-E_B(K))
 -\Theta(\omega-E_{\rm th}(K))\delta A_{\rm scatt}({\bf K},\omega),
 \label{eq:rel_spectral_ansatz}
\end{align}
Here $\mathcal B>0$ corresponds to the binding energy of a baryon, $\delta A_{\rm scatt}>0$ is the scattering-state contribution.  The frequency
sum rule is given by
\begin{equation}
 \int_{-\infty}^{\infty}d\omega\,\Delta A_B({\bf K},\omega)=0,
 \qquad
 \int_{E_{\rm th}(K)}^\infty d\omega\,
 \delta A_{\rm scatt}({\bf K},\omega)=1.
 \label{eq:rel_full_sumrule}
\end{equation}
At $T=0$, the spectral integral in Eq.~\eqref{eq:rel_phase_density} is restricted to $\omega<\mu_B$.
Writing $\mu_B^{\rm kin}=N_c(\mu-m_Q)+\mathcal B$, the interval
$m_Q-\mathcal B/N_c<\mu<m_Q$ contains occupied baryon poles but no
quark sea or occupied continuum, giving an ordinary baryon Fermi gas.

For $\mu\geq m_Q$, both the quark term and the negative continuum term
become active.  
We introduce the quark and baryon Fermi momenta
 $k_{FQ}=\sqrt{\mu^2-m_Q^2}$ and
 $k_{FB}=\sqrt{2M_B\mu_B^{\rm kin}}$, respectively.
Writing the continuum energy as $\omega=E_{\rm th}(K)+\epsilon$, the occupied weight of the scattering states is
\begin{equation}
 I(K)=\Theta(X_K)\int_0^{X_K}d\epsilon\,
 \delta A_{\rm scatt}({\bf K},E_{\rm th}(K)+\epsilon),
 \label{eq:rel_continuum_weight}
\end{equation}
where $X_K=\mu_B-E_{\rm th}(K)$.
Equation~\eqref{eq:rel_continuum_weight} is defined to vanish for $X_K\leq0$.  
Hence it contributes
only for $K<K_0=\sqrt{2M_BN_c(\mu-m_Q)}$, and the density becomes
\begin{equation}
 n_B=\frac{g}{6\pi^2}(k_{FQ}^3+k_{FB}^3)
 -g\sum_{\bf K}I(K).
 \label{eq:rel_density_before_saturation}
\end{equation}
At this stage, the shell is depleted at $K<K_0$, but this state does not need
to be empty: the full-frequency sum rule does not determine how much
continuum weight lies below $\mu_B$.

The additional approximation used in Ref.~\cite{Tajima2026} is that,
for $K<K_0$, the available continuum weight is already exhausted within
the occupied energy interval:
\begin{equation}
 \int_{-\infty}^{\mu_B}d\omega\,
 \Delta A_B({\bf K},\omega)\simeq0
 \quad(K<K_0),\qquad
 I(K)\simeq\Theta(K_0-K).
 \label{eq:rel_occupied_saturation}
\end{equation}
Because $k_{FB}>K_0$, this cancels the occupied pole inside $K_0$, leaving

\begin{align}
 f_B(K)&\simeq\Theta(k_{FB}-K)-\Theta(K_0-K)
 =\Theta(K-K_0)\Theta(k_{FB}-K).
 \label{eq:rel_shell_distribution}
\end{align}
Thus the inner boundary follows from the occupied continuum threshold,
and the outer boundary follows from the baryon pole.

Performing the momentum integrals in
Eq.~(\ref{eq:rel_density_before_saturation}),
we obtain
\begin{align}
 n_B&\simeq\frac{2}{3\pi^2}
 \left[k_{FQ}^3+k_{FB}^3-(k_{FB}-\Delta_B)^3\right],
 \label{eq:rel_MR_density}
\end{align}
where
\begin{align}
 \Delta_B&=k_{FB}-K_0
 \equiv\sqrt{2M_B[N_c(\mu-m_Q)+\mathcal B]}
 -\sqrt{2M_BN_c(\mu-m_Q)}
 \label{eq:rel_shell_width}
\end{align}
is the baryonic momentum-shell width.
Equation~(\ref{eq:rel_MR_density}) consists of the quark Fermi sea plus baryon-shell
counting formula of McLerran and Reddy~\cite{McLerranReddy2019}, recovered
as in Ref.~\cite{Tajima2026}.  In this construction, $\Delta_B$
arises from subtracting the occupied scattering continuum from the
baryonic pole contribution.

The original McLerran--Reddy model additionally relates the momenta by
$k_{FQ}=(k_{FB}-\Delta_B)/N_c=K_0/N_c$ and specifies a phenomenological
density dependence of the shell width~\cite{McLerranReddy2019}.
With the mixed relativistic-quark and nonrelativistic-baryon dispersions
used here, $K_0\simeq N_c k_{FQ}$ holds near the quark threshold when
$M_B\simeq N_c m_Q$ and $\mu-m_Q\ll m_Q$.
The phase-shift construction therefore establishes the
microscopic structure and mechanism for the baryon shell, while its detailed
kinematics and width depend on the assumed spectra.

\section{Summary and outlook}
\label{sec:summary}

Tripling fluctuations offer a microscopic mechanism connecting three-fermion
bound states to degenerate fermionic matter across a continuous three-body crossover.
The phase-shift representation identifies the redistribution of spectral weight
between a baryonic pole and the scattering continuum as the common origin of
both a baryonic momentum shell and a peaked speed of sound.  This perspective
provides a bridge between ultracold-atom many-body theory and phenomenological
models of quarkyonic matter.
Its next challenge is to retain this
mechanism in a quantitatively controlled equation of state incorporating the
symmetries, composition, and interactions of dense QCD matter.

\section*{Acknowledgments}
The author thanks T.~M.~Doi, K.~Iida, T.~Kojo, H.~Liang, and S.~Tsutsui for useful discussions during the studies of Refs.~\cite{Tajima2022,Tajima2025,Tajima2026}.
This work was supported by 
Japan Society for the Promotion of Science (JSPS) Grants-in-Aid for Scientific Research (KAKENHI) Grant Nos.~JP22K13981, JP23K22429, and JP26K07063.

\section*{Declaration of generative AI and AI-assisted technologies in the
manuscript preparation process}

During the preparation of this work, the author used ChatGPT (OpenAI) to
assist with literature organization, manuscript structure, and language
improvement.  After using this tool, the author reviewed, verified, and
edited the content as needed and takes full responsibility for the content of
the publication.

\bibliographystyle{elsarticle-num}
\bibliography{reference}

\end{document}